\documentclass[namedreferences]{solarphysics}

\usepackage[hyperref,optionalrh,showbiblabels]{spr-sola-addons}

\usepackage[utf8]{inputenc}
\usepackage{graphicx}
\usepackage{textcomp}
\usepackage{color}
\usepackage{breakurl}

\newcommand{\aap}{    {\it Astron. Astrophys.}}

\newcommand{\aapr}{   {\it Astron. Astrophys. Rev.}}

\newcommand{\apj}{    {\it Astrophys. J.}}
\newcommand{\apjl}{   {\it Astrophys. J. Lett.}}

\newcommand{\grl}{    {\it Geophys. Res. Lett.}}

\newcommand{\jgr}{    {\it J. Geophys. Res.}}
\newcommand{\mnras}{  {\it Mon. Not. Roy. Astron. Soc.}}

\newcommand{\prl}{    {\it Physical Review Letters}}
\newcommand{\solphys}{{\it Solar Phys.}}
 
\newcommand{\ssr}{    {\it Space Sci. Rev.}} 
\chardef\us=`\_

\begin{document}

\begin{article}
\begin{opening}

\title{Two-stage evolution of an extended C-class eruptive flaring activity from sigmoid active region NOAA 12734: \textit{SDO} and Udaipur-CALLISTO observations}

\author[addressref=aff1,corref,email={bhuwan@prl.res.in}]{\inits{B.}\fnm{Bhuwan}~\lnm{Joshi}\orcid{https://orcid.org/0000-0001-5042-2170}}
\author[addressref={aff1,aff2}]{\inits{PKM}\fnm{Prabir K.}~\lnm{Mitra}\orcid{https://orcid.org/0000-0002-0341-7886}}
\author[addressref=aff1]{\inits{}\fnm{R.}~\lnm{Bhattacharyya}\orcid{https://orcid.org/0000-0003-4522-5070}}
\author[addressref=aff1]{\inits{}\fnm{Kushagra}~\lnm{Upadhyay}\orcid{https://orcid.org/0000-0001-6261-2756}}
\author[addressref=aff3]{\inits{}\fnm{Divya}~\lnm{Oberoi}\orcid{https://orcid.org/0000-0002-4768-9058}}
\author[addressref=aff4]{\inits{}\fnm{K. Sasikumar}~\lnm{Raja}\orcid{https://orcid.org/0000-0002-1192-1804}}
\author[addressref=aff5]{\inits{}\fnm{Christian}~\lnm{Monstein}\orcid{https://orcid.org/0000-0002-3178-363X}}

\address[id=aff1]{Udaipur Solar Observatory, Physical Research Laboratory, Udaipur 313001, India}
\address[id=aff2]{Department of Physics, Gujarat University, Ahmedabad 380 009, India}
\address[id=aff3]{National Centre for Radio Astrophysics, Tata Institute of Fundamental Research, Pune 411007, India}
\address[id=aff4]{Indian Institute of Astrophysics, Bangalore 560034, India}
\address[id=aff5]{Istituto Ricerche Solari Locarno (IRSOL), 6605 Locarno Monti, Switzerland}

\runningauthor{Bhuwan Joshi et al.}
\runningtitle{Extended C-class flaring activity}

\begin{abstract}
In this article, we present a multi-wavelength investigation of a C-class flaring activity that occurred in the active region NOAA 12734 on 8 March 2019. The investigation utilises data from the \textit {Atmospheric Imaging Assembly} (AIA) and the \textit {Helioseismic Magnetic Imager (HMI)} on board the \textit {Solar Dynamics Observatory} (SDO) and the Udaipur-CALLISTO solar radio spectrograph of the Physical Research Laboratory. This low intensity C1.3 event is characterised by typical features of a long duration event (LDE), viz. extended flare arcade, large-scale two-ribbon structures and twin coronal dimmings. The eruptive event occurred in a coronal sigmoid and displayed two distinct stages of energy release, manifested in terms of temporal and spatial evolution. The formation of twin dimming regions are consistent with the eruption of a large flux rope with footpoints lying in the western and eastern edges of the coronal sigmoid. The metric radio observations obtained from Udaipur-CALLISTO reveals a broad-band ($\approx$50-180 MHz), stationary plasma emission for $\approx$7 min during the second stage of the flaring activity that resemble a type IV radio burst. A type III decametre-hectometre radio bursts with starting frequency of $\approx$2.5 MHz precedes the stationary type IV burst observed by Udaipur-CALLISTO by $\approx$5 min. The synthesis of multi-wavelength observations and Non-Linear Force Free Field (NLFFF) coronal modelling together with magnetic decay index analysis suggests that the sigmoid flux rope underwent a zipping-like uprooting from its western to eastern footpoints in response to the overlying asymmetric magnetic field confinement.  The asymmetrical eruption of the flux rope also accounts for the observed large-scale structures \textit{viz} apparent eastward shift of flare ribbons and post flare loops along the polarity inversion line (PIL), and provides an evidence for lateral progression of magnetic reconnection site as the eruption proceeds.
\end{abstract}

\end{opening}

\section{Introduction}
A solar flare is characterised by a sudden catastrophic release of energy from localised regions of the solar atmosphere \citep{Benz2017}. They are often accompanied with the eruption of plasma and energetic particles and, therefore, extensively studied in view of the origin of space weather manifestations. The temporal and spatial scales of a flare often represent the scale of energy release that ranges from $<$10$^{27}$~erg to $>$10$^{32}$~erg. Magnetic reconnection has been recognised as the basic process that converts the stored magnetic energy of active region into heating of plasma, acceleration of particles, and kinematic motion of coronal plasma \citep[see review by][]{Priest2002}. 

With the advent of space-borne flare observations in soft X-ray (SXR) energies in 1970s, it was recognised that flares, in general, follow two morphologically distinct classes: confined and eruptive events \citep{Pallavivini1977}. The confined flares undergo brightening in compact loop structures with little large-scale motion. Topological models consider the energy release in confined flares within a single static loop and, therefore, these events are also referred to as single-loop or compact flares. Being non-eruptive, the large-scale magnetic configuration of the active region corona is largely preserved in confined flares \citep[\textit{e.g.}][]{Kushwaha2014, Ning2018}. The second category comprises the long duration events (LDE) which are eruptive in nature. They are mostly associated with coronal mass ejections (CMEs) and exhibit formation of an arcade of coronal loops during the prolonged energy release. The observations of eruptive flare in chromospheric imaging channels ({\it e.g.} H$\alpha$ and EUV 304~\AA) readily show formation of conjugate flare ribbons on opposite magnetic polarity regions and, therefore, eruptive flares are also called two-ribbon flares \citep[see reviews by][]{Vrsnak2003,Fletcher2011}. Despite numerous observations, we still do not have a clear understanding on how the changes in the magnetic flux through the photosphere constrain the filament activation, leading to LDEs. Once a flare is triggered, regardless of its confined or eruptive nature, depending upon the complex topology of field lines at the core and overlying regions, we observe multi-wavelength emissions which enhance our understanding about underlying radiative processes \citep[\textit{e.g.}][]{Kushwaha2015,JoshiNC2019}. Therefore, the multi-wavelength case studies of LDEs form an important topic of research to understand the physics of large-scale magnetic reconnection and initiation mechanisms of CMEs.

Solar flares are powered by excess magnetic energy stored in active region in the form of stressed magnetic field. Observations of coronal sigmoids -- the S-shaped coronal features frequently visible in SXR and EUV images \citep{Manoharan1996,Rust1996,Joshi2017} -- provide evidence for the sites of magnetic energy storage to power solar eruptions. Sigmoids are highly twisted, large-scale loops that sometime encompass almost full active region area and show high association with eruptive flares \citep{Canfield1999}. Indeed, multi-wavelength analysis of sigmoidal active regions and related coronal magnetic field modelling provide evidence for the existence of highly non-potential magnetic fields configuration in the pre-CME corona in the form of hot EUV coronal channel \citep[\textit{e.g.}][]{Cheng2014a, Mitra2019}, H$\alpha$ filaments \citep[\textit{e.g.}][]{Pevtsov2002, Sahu2020} and magnetic flux ropes \citep[\textit{e.g.}][]{Inoue2018, Mitra2018, Mitra2021}. Imperatively, temporal and spatial investigations of energy release processes during solar eruptive flares and its comparison with complex overlying magnetic field structures are necessary to probe the flare emission produced at different heights of solar atmosphere as a result of progression of magnetic reconnection in the corona.

In this paper, we present the multi-wavelength analysis of C-class flaring activity that occurred on 8 March 2019 (SOL2019-03-08T03:19) in active region (AR) NOAA 12734.  
An important characteristic of the event lies in its two-stage evolution both in terms of temporal and spatial aspects. In Section~\ref{sec:data}, we briefly document the sources of observational data. We study the structure and evolution of the source AR NOAA 12734 in Section~\ref{sec:AR12734}. The different temporal and spatial aspects of the energy release are explored in Sections~\ref{sec:goes-lc}--\ref{sec:AIA} where we mainly analyse data from the \textit{Solar Dynamics Observatory} \cite[\textit{SDO};][]{Pesnell2012}. In Section \ref{sec:callisto}, we briefly discuss CALLISTO radio spectrograph at Udaipur Solar Observatory (USO), Physical Research Laboratory (PRL) and present observations of flare-associated radio burst. The coronal magnetic configuration of the AR is investigated in Section~\ref{sec:extrapolation}. We discuss and summarise our observational results in Section~\ref{sec:discussion}.

\section{Observational data}
\label{sec:data}
To probe the energy release processes simultaneously at different temperature regions of solar atmosphere during the C1.3 flare, we analyse the high temporal and spatial resolution images at various (E)UV channels obtained from the \textit{Atmospheric Imaging Assembly} \citep[AIA;][]{Lemen2012} on board the \textit{SDO}. The images taken from the \textit{Helioseismic Magnetic Imager} \citep[HMI;][]{Schou2012} on board the \textit{SDO} are used to understand the basic photospheric structure of the active region that include strong field regions associated with sunspots and magnetically weaker plage regions. The CALLISTO radio spectrograph system at USO/PRL (known as Udaipur-CALLISTO; see Section \ref{sec:callisto} for details), observed subtle yet distinct signatures of the flare in the frequency range of $\approx$50--180~MHz. Notably, the radio burst associated with C-class flaring activity on 8 March 2019 happens to be the first solar event observed by the Udaipur-CALLISTO system since its commencement in October 2018. To study the flare associated signatures at low frequencies, below the ionospheric cut-off $\lessapprox$15~MHz, we analyse dynamic radio spectrum obtained from Radio and Plasma Wave Experiment \citep[WAVES;][]{Bougeret1995} on board Wind spacecraft.

We carry out coronal magnetic field extrapolation by employing the optimisation based Non-Linear Force Free Field (NLFFF) extrapolation method \citep{Wiegelmann2010, Wiegelmann2012}, using a pre-flare photospheric vector magnetograms from the ``hmi.sharp$\_$cea$\_$720s" series of HMI/\textit{SDO} as boundary condition. A summary of the optimisation based NLFFF technique is provided in \citet{Mitra2020b}. Extrapolation was done in a Cartesian box of dimensions 342$\times$195$\times$195 pixels with pixel resolution 0.''5 pixel$^{-1}$ which corresponds to a physical volume\footnote{At a distance of 1 AU, an angular diameter of 1 arcsec represents $\approx$0.725 Mm on the Sun. As a standard assumption, this conversion factor is uniformly applied across the whole magnetogram.} having dimensions $\approx$124$\times$71$\times$71 Mm$^3$. For visualising the extrapolated coronal field lines, we have used Visualization and Analysis Platform for Ocean, Atmosphere, and Solar Researchers \citep[VAPOR;][]{Clyne2007} software.

\section{Structure of the source region NOAA 12734}
\label{sec:AR12734}

\begin{figure*}
\centering
\includegraphics[width=\textwidth]{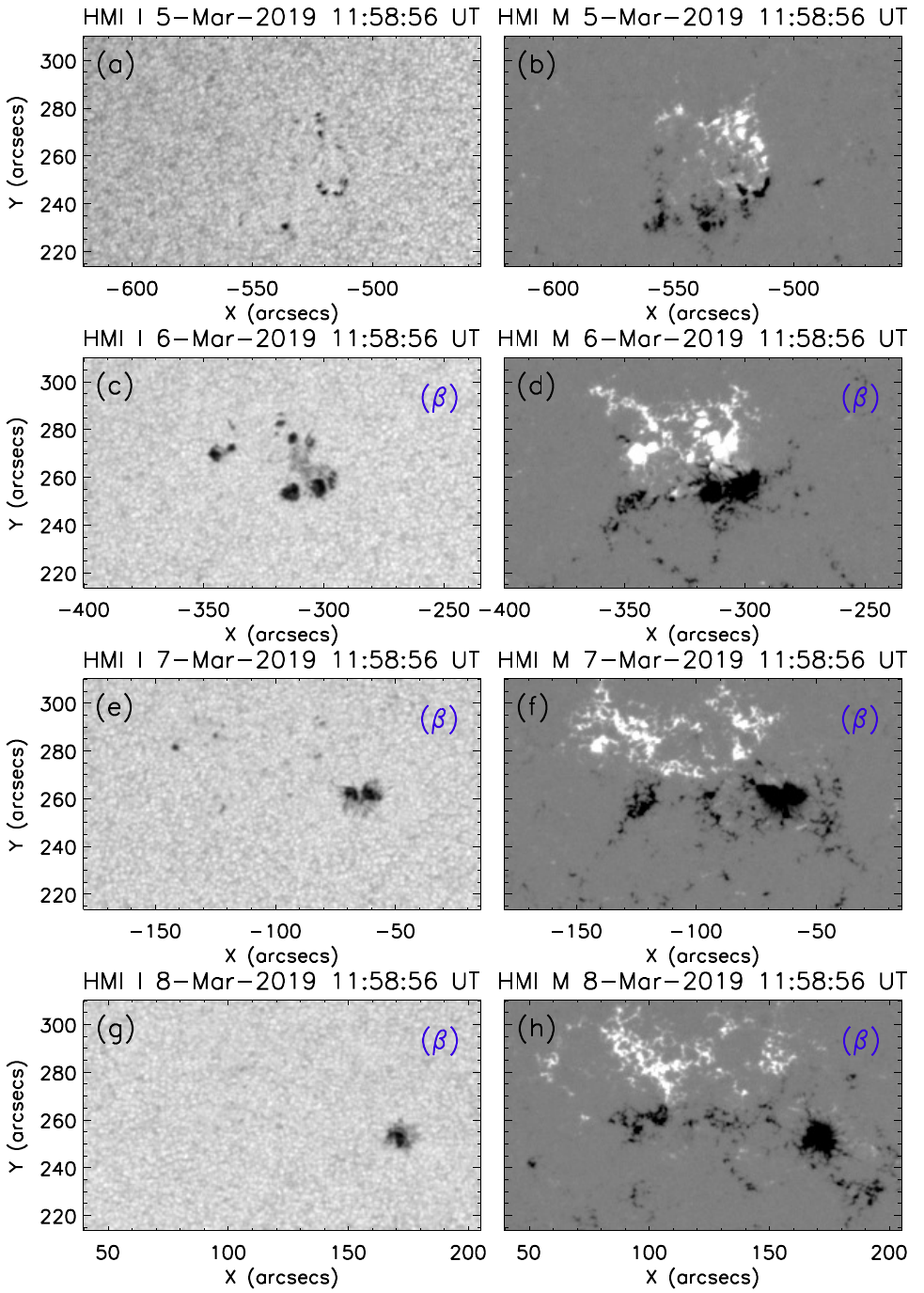}
\caption{HMI intensity images (left panels) and Line-of-Sight (LOS) magnetograms (right panels) showing the evolution of AR NOAA 12734 during 5-8 March 2020. The AR evolved into a $\beta$-type configuration on 6 March and remained so until 8 March.}
\label{Fig_AR_evolution}
\end{figure*}

The first signatures of NOAA 12734 appeared in HMI magnetogram images on March 4 with the emergence of dispersed magnetic flux regions of positive and negative polarities. There was no manifestation of sunspots in HMI intensity images on this date. In Figure~\ref{Fig_AR_evolution}, we show the evolution of AR with intensity images and HMI magnetograms. We note that, the AR evolved rapidly on the subsequent days (March 5--6) with clear appearance of several small sized sunspots in leading and following parts of the AR with leading sunspot group being much more prominent than the later (Figure~\ref{Fig_AR_evolution}). After March 6, the AR sunspots started to decay. Notably, during the decay phase, photospheric magnetic flux of the AR underwent dispersion and extended in a larger region (Figures~\ref{Fig_AR_evolution}e--h). 

\begin{figure*}
\centering
\includegraphics[width=10cm]{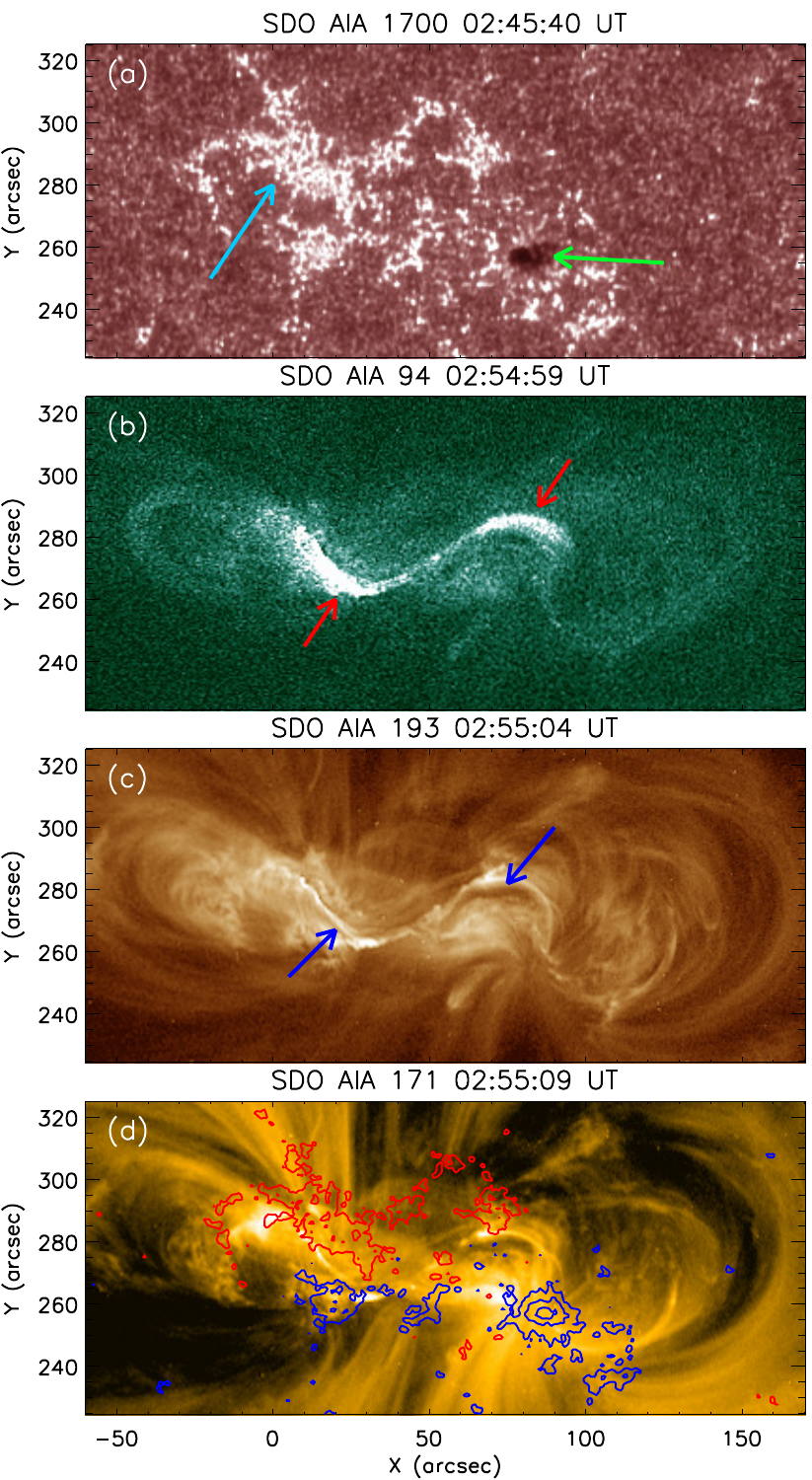}
\caption{Multi-wavelength view of AR NOAA 12734 showing the circumstances where C1.3 long duration flare occurred on 8 March 2019 between 03:00-05:00~UT. (\textbf{a}): UV (1700~\AA) image showing sunspots (green arrow) and plage region (sky of the AR NOAA 12734. (\textbf{b}) and (\textbf{c}): AIA~193~{\AA} and 94~{\AA}~images of the AR prior to the event, showing hot coronal loops that are twisted to form an overall S-shape \textit{i.e.} a sigmoid structure (see arrows in \textbf{b}). A filament channel is also seen in the AIA~193~\AA~image which is marked by arrows. (\textbf{d}): AIA~171~\AA~pre-flare images showing the coronal loops in and around the sigmoid. Red and blue contours represent HMI LOS magnetograms with positive and negative polarities, respectively.}
\label{Fig_AR}
\end{figure*}

On March 8 when the reported activity occurred, only the leading sunspot was visible in intensity images. The comparison of HMI magnetogram (Figure~\ref{Fig_AR_evolution}g) with intensity  (Figure~\ref{Fig_AR_evolution}h) suggests that the leading sunspot region is of negative polarity while there are sunspot-less dispersed flux regions of mix polarities in the following part of the active region. The structure of the active region at multiple heights and temperatures is further investigated in Figure~\ref{Fig_AR}. The AIA~1700~\AA~image (log(T)=3.7) shows the leading sunspot while plage-like brightenings can be noticed away from the sunspot but within the active region (Figure~\ref{Fig_AR}a). In panels b and c of Figure~\ref{Fig_AR}, we show the AIA~94~\AA~(log($T$)=6.8) and AIA~193~\AA~(log($T$) = 6.2, 7.3) images, respectively, which reveal bright, hot coronal loops that are twisted and form an overall S-shaped structure, \textit{i.e.} a coronal sigmoid. We also find dark thread-like features in 193~\AA~images (marked by arrows in Figure~\ref{Fig_AR}c) which points toward the presence of a filament channel. The co-temporal AIA 171~\AA~image (log($T$) = 5.8; Figure~\ref{Fig_AR}d) together with photospheric LOC magetogram (as contours) shows coronal loop system and corresponding magnetic polarities near their footpoint locations.

\section{GOES light curves: two-step energy release process}
\label{sec:goes-lc}
\begin{figure*}
\centering
\includegraphics[width=\textwidth]{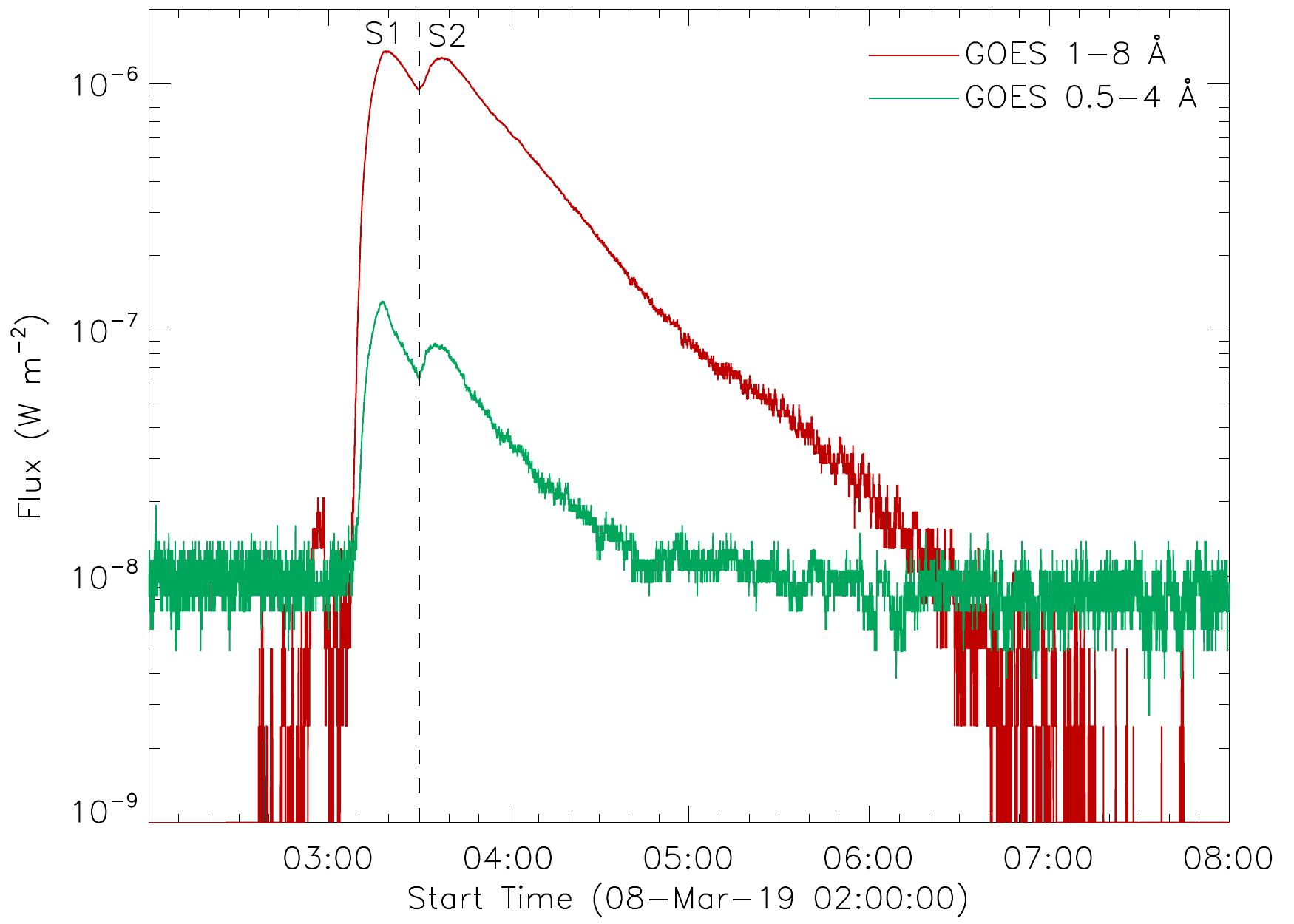}
\caption{GOES light curves showing the evolution of an extended C-class flaring activity in active NOAA 12734. The red and green lines indicate X-ray flux in 1--8~\AA~and 0.5--4~\AA~wavelength bands which correspond to disk-integrated X-ray emission in 1.5--12.5 keV and 3--25 keV energy range, respectively. GOES profiles reveal two distinct episodes of energy release (marked as S1 and S2) that peak at 03:19 and 03:38~UT, respectively, implying two-step process of energy release.}
\label{Fig_goes}
\end{figure*}

In Figure~\ref{Fig_goes}, we provide the time variation of soft X-ray (SXR) flux in GOES channels of 1--8~\AA~and 0.5--8~\AA~between 02:00 UT and 08:00 UT. We note that in 1--8~\AA~channel, the flux before $\approx$03:00 UT is extremely low and in GOES terminology it is much below the level of an $\approx$A0 class flare. Starting from $\approx$03:00 UT, we find an abrupt rise of flux in both the channels that imply the rise phase of the flare. 
The peak flux in 1--8~\AA~channel indicates a flare of class C1.3 with peak at $\approx$03:19 UT after which the flux decayed in both the channels until $\approx$03:30 UT. It is remarkable to note the development of a secondary stage of emission from $\approx$03:30 UT onward that again exhibit distinct peak at $\approx$03:38 UT in 1--8~\AA~light curve. Notably, the higher energy profile of 0.5--4~\AA~also presents similar time evolution. The 1--8~\AA~profile reaches the pre-flare background level after $\approx$06:30 UT implying a very prolonged energy release phase of $\approx$3.5 hours. The 0.5--8~\AA~profile, however, shows the arrival of pre-flare emission level after $\approx$04:40~UT. 

\section{(E)UV imaging observations}
\label{sec:AIA}

\subsection{Sequential brightenings of the coronal loop system}
\begin{figure*}
\centering
\includegraphics[width=\textwidth]{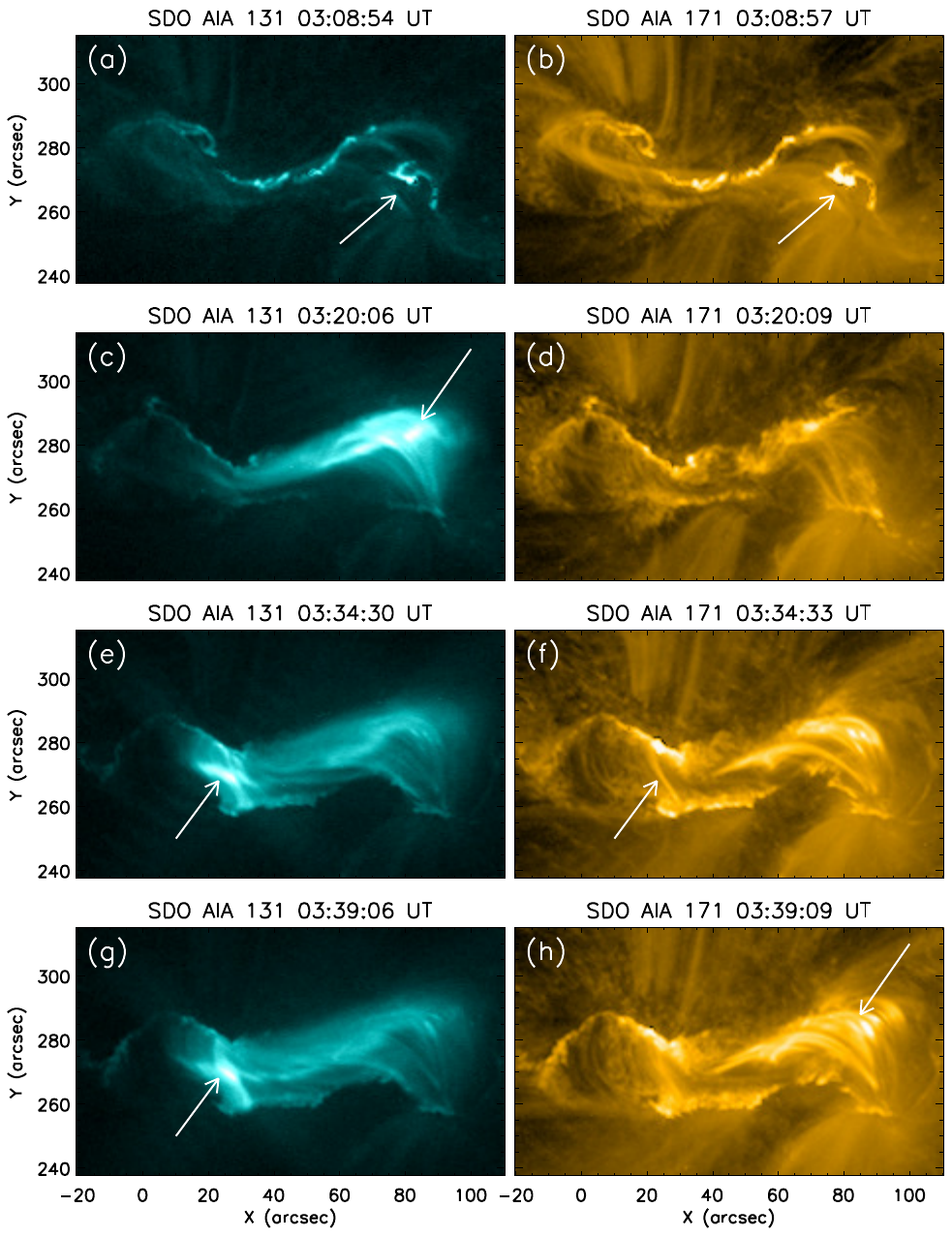}
\caption{Series of AIA images in AIA 131 and 171~\AA~showing the evolution of the long duration event. We recall that GOES time profile (Figure~\ref{Fig_goes}) shows two stages of energy release: S1 and S2 peaking at 03:19 and 03:38 UT, respectively. Sequence of images clearly reveal that change of coronal energy release site from western to eastern part implying two distinct events of coronal magnetic reconnection. An animation of this figure is provided in the supplementary materials.} 
\label{Fig_fl_131_171}
\end{figure*}

We use AIA observations to probe the spatial characteristics of the primary and secondary energy release stages, as noted in GOES profiles (see Figure~\ref{Fig_goes} and Section~\ref{sec:goes-lc}) using (E)UV imaging observations. For the purpose, in Figure~\ref{Fig_fl_131_171}, we show flaring activity in two representative AIA channels, 131~\AA~(log($T$) = 5.6, 7.0) and 171~\AA~(log($T$) = 5.8). Notably, AIA 131~\AA~channel also contains contribution of hot plasma and, therefore, suitably images the emission from the flaring region.
The sequence of images clearly reveals that the flare triggered at the western part of the AR (indicated by arrow in Figures~\ref{Fig_fl_131_171}a and b). Further, during the first stage of the flaring activity, the emission dominated at the western segment of the coronal sigmoid. During his stage, brightness at the eastern part of the active region slowly builds up. The AIA 131~\AA~images near the peak of the first stage (which is also the overall peak of the flaring activity) reveal the formation of a set of bright coronal loops (indicated by an arrow in Figure~\ref{Fig_fl_131_171}c). In 171~\AA~observations, these loops appear much fainter (Figure~\ref{Fig_fl_131_171}d). 

During the second stage of the flare, we clearly see energy release (\textit{i.e.} flare brightening) and subsequent morphological changes in 131~\AA~images that occurred in the eastern part of the active region (indicated by arrow in Figures~\ref{Fig_fl_131_171}e--g). Flare associated brightenings can also be noticed in 171~\AA~images, albeit less prominent in comparison to the hot 131~\AA~images. Interestingly, 171~\AA~images of the AR during this time display a dense post-flare arcade at the location of first stage of energy release (indicated by the arrow in Figure~\ref{Fig_fl_131_171}h).

\subsection{Evolution of post-flare loop arcade and flare ribbons} 
\label{sec:loops_ribbons}
\begin{figure*}
\centering
\includegraphics[width=\textwidth]{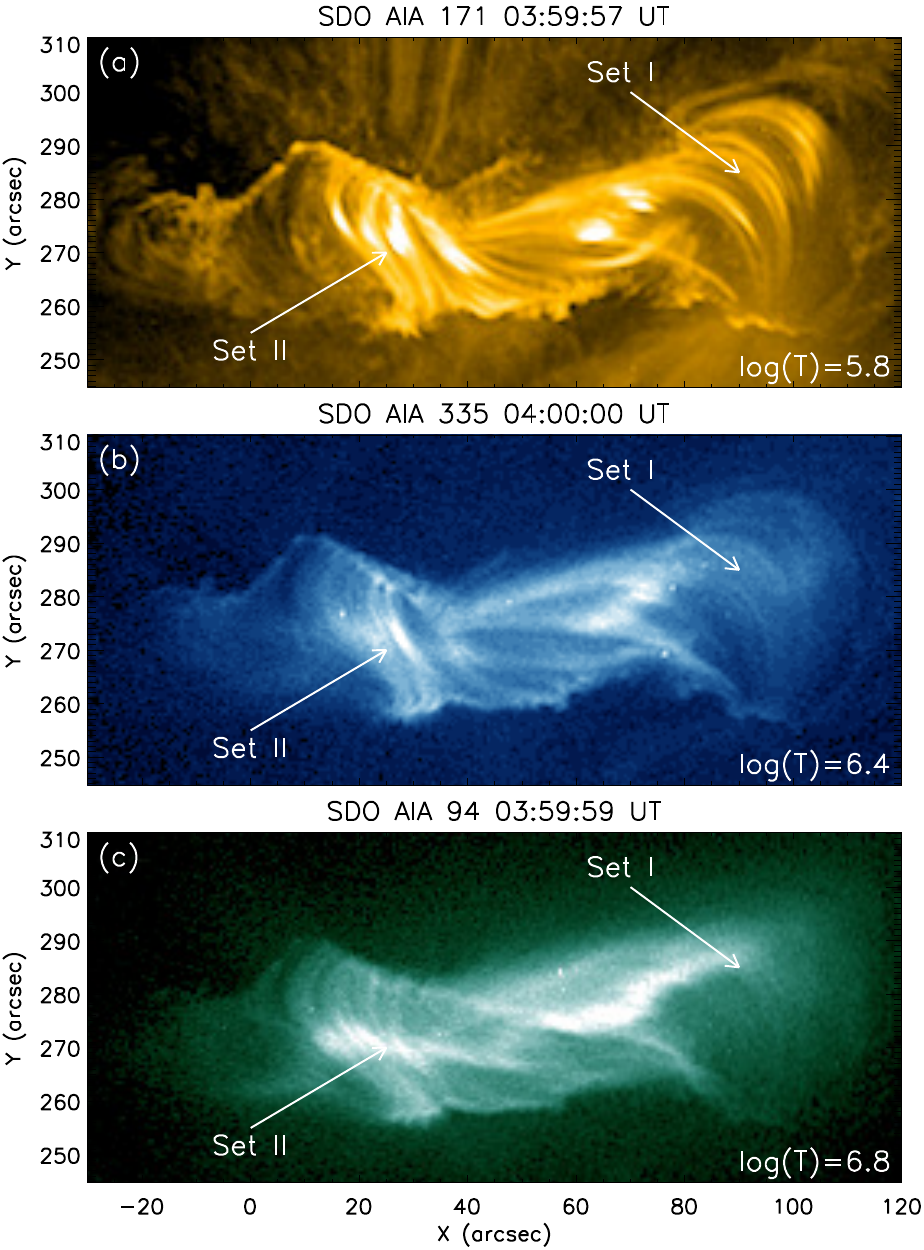}
\caption{AIA images at multiple channels showing the captivating post-flare arcades during the prolonged decay phase of the C1.3 flare. We recognize two sets of loop systems, marked as set I and II, which are sequentially formed during the stage I (S1) and stage II (S2) of the long duration event.}
\label{Fig_arcade}
\end{figure*}

The EUV images exhibit an apparent eastward shift in the site of energy release during the long duration event.
However, with the passage of time, we observe the formation of dense system of post-flare loops at both the locations which is shown in Figure~\ref{Fig_arcade}. We also note that in the 171~\AA~images, the loop arcade presents a much structured emission where  multiple loops can be clearly resolved at the location of energy release during first and second stages (corresponding loop systems are denoted as set I and set II in Figure~\ref{Fig_arcade}). In 94~\AA~channel, the coronal region appears very diffused, indicating that the loops are filled with very hot plasma (log($T$) = 6.8).

\begin{figure*}
\centering
\includegraphics[width=\textwidth]{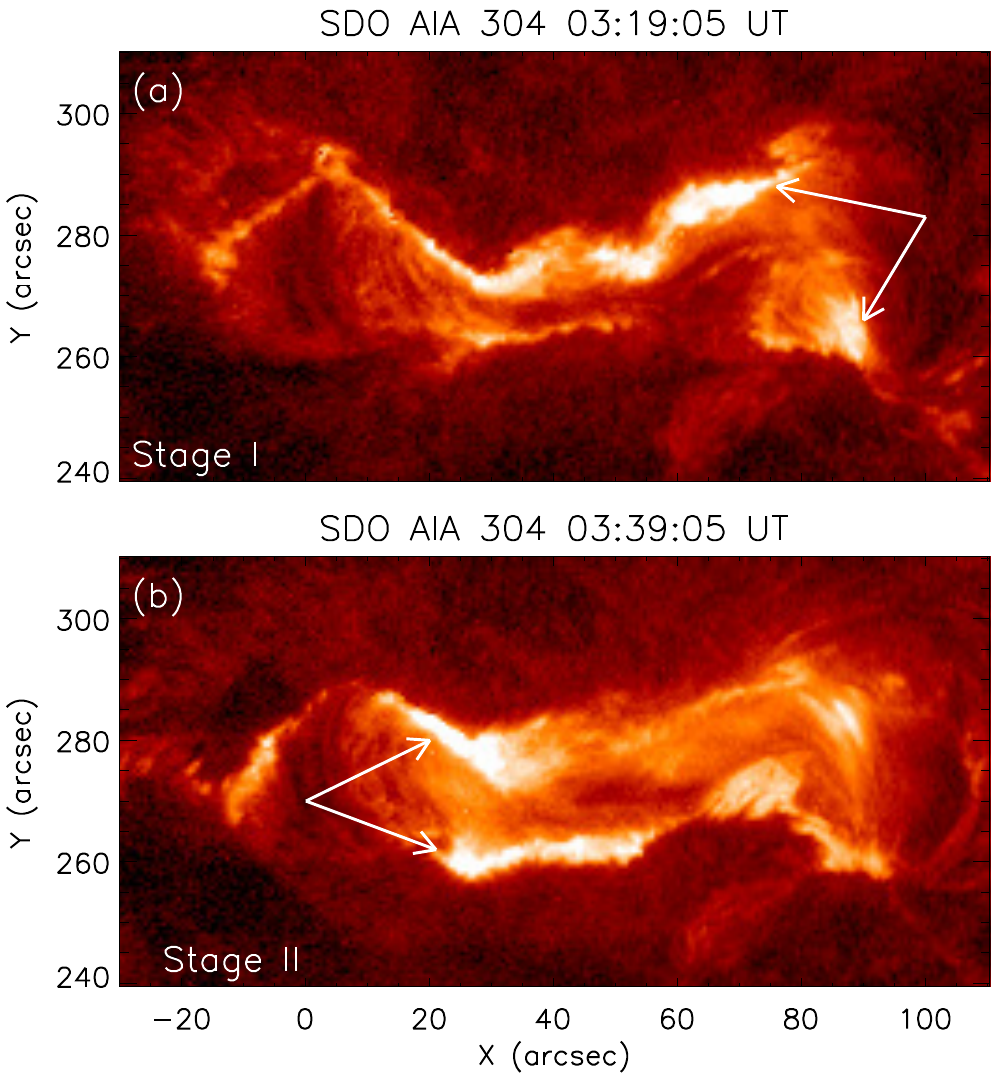}
\caption{Representative AIA 304~\AA~images showing the sequential development of flare ribbons with the progression of coronal energy release. During stage I of the event, western part of the ribbons exhibited enhanced brightnening. Notably, during stage I, the northern ribbon presented an extended structure while the southern ribbon is spatially localised. In stage II, the brigntening progressed toward the eastern part. An animation of this figure is provided in the supplementary materials.}
\label{Fig_ribbons}
\end{figure*}

Complimentary to the observations of sequential energy release during the first and second stages and corresponding coronal activities, we find spatial variations in the brightness of transition region and chromospheric layers. Figure~\ref{Fig_ribbons} presents AIA~304~\AA\ images of the flaring region during the time of first and second peaks. We find prominent two-ribbon brightenings near the western part of active region during the first stage (indicated by arrows in Figure~\ref{Fig_ribbons}a). Among the two ribbons, the northern one was spatially extended while the southern one exhibited localised brightening. Subsequently, the ribbons expanded laterally as the second stage approached. The peak of the second phase of the flare witnessed two well-developed flare ribbons toward the eastern part of coronal sigmoid (indicated by arrows in Figure~\ref{Fig_ribbons}b). Notably, contrary to the ribbons of the first stage, the ribbons of second stage were comparable in length.

\subsection{Coronal dimmings and subsequent CME}
\begin{figure*}
\centering
\includegraphics[width=\textwidth]{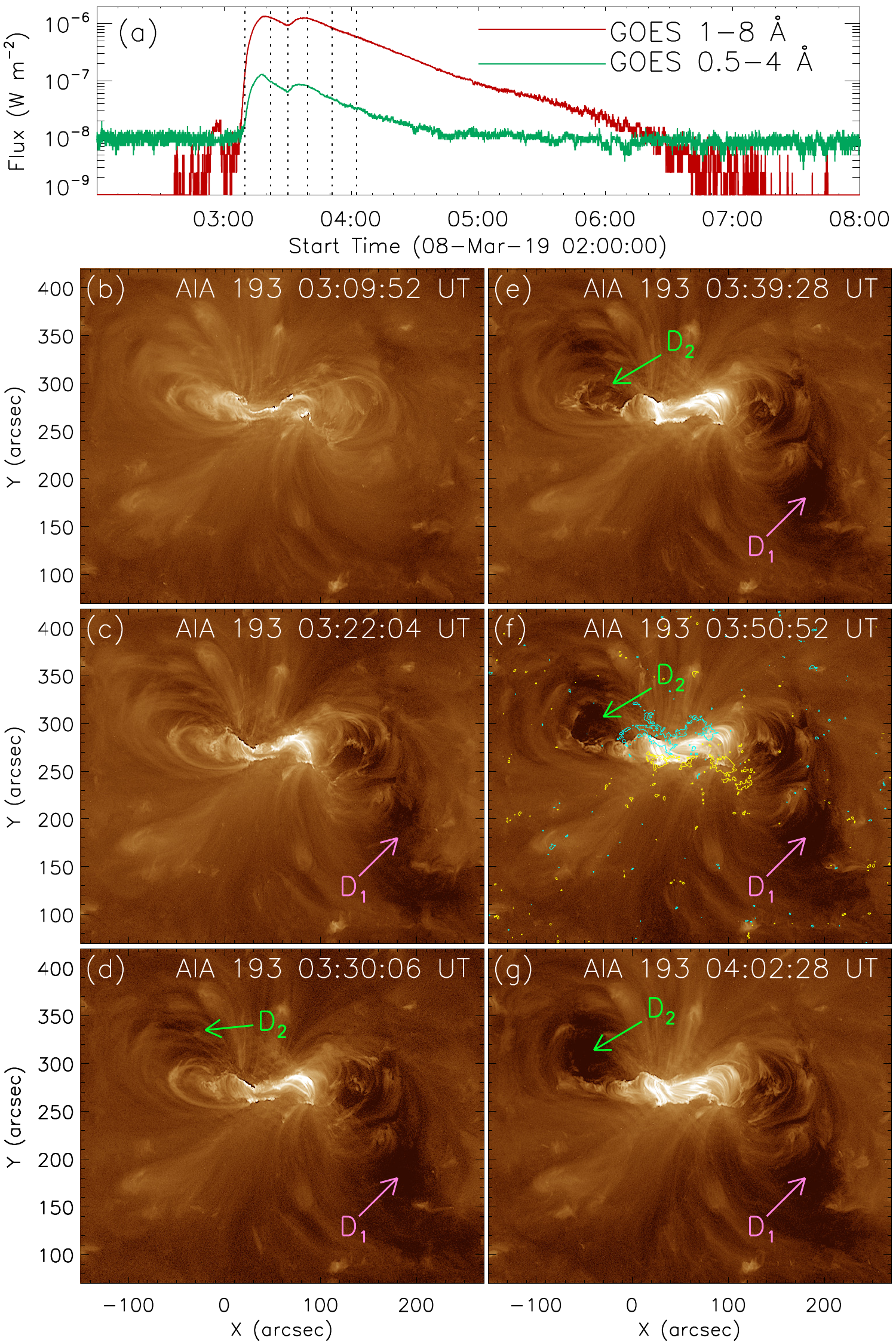}
\caption{Series of AIA images in AIA 193~\AA~showing the sequential development of coronal dimming regions D$_{1}$ and D$_{2}$ following the first and second stages of the flare, respectively (\textbf{b})-(\textbf{g}). For a clear understanding of the development of the dimming regions with the flare evolution, we present GOES light curve in (\textbf{a}). The vertical lines in GOES light curves represent the timings of AIA~131~\AA~images shown in panels (\textbf {b})-(\textbf{g}). Co-temporal HMI LOS magnetogram is overplotted in (\textbf{f}). Sky and yellow contours represent positive and negative magnetic polarities, respectively.} 
\label{Fig_coronal_dimming}
\end{figure*}

In Figure~\ref{Fig_coronal_dimming}, we show the evolution of the eruptive activity in the AIA 193 \AA~images (log($T$) = 6.2, 7.3). To understand the chronological development of the coronal features, we also provide the GOES light curve in panel a where the vertical dashed lines represent the timings of representative EUV images in panels b--g. We can readily note the development of regions of reduced EUV brightenings to the southwest (marked as D1 region) and northeast (marked as D2 region) of the coronal sigmoid. Reduced regions of EUV intensity are refereed to as the coronal dimmings \citep{Hudson1996,Mandrini2007}. The D1 region appears first, around the first peak, while the D2 region starts developing in the later stages, when the first peak decays but before the beginning of the second stage of SXR emission. In Figure~\ref{Fig_coronal_dimming}f, we overplot photospheric magnetograms on the EUV image which suggest that the twin dimming regions are formed in opposite magnetic polarities; this spatial association relates that the opposite polarity feet of the fluxrope terminate at the location of dimming regions in the pre-eruption stage \citep{Webb2000}.

\begin{figure*}
\centering
\includegraphics[width=0.5\textwidth]{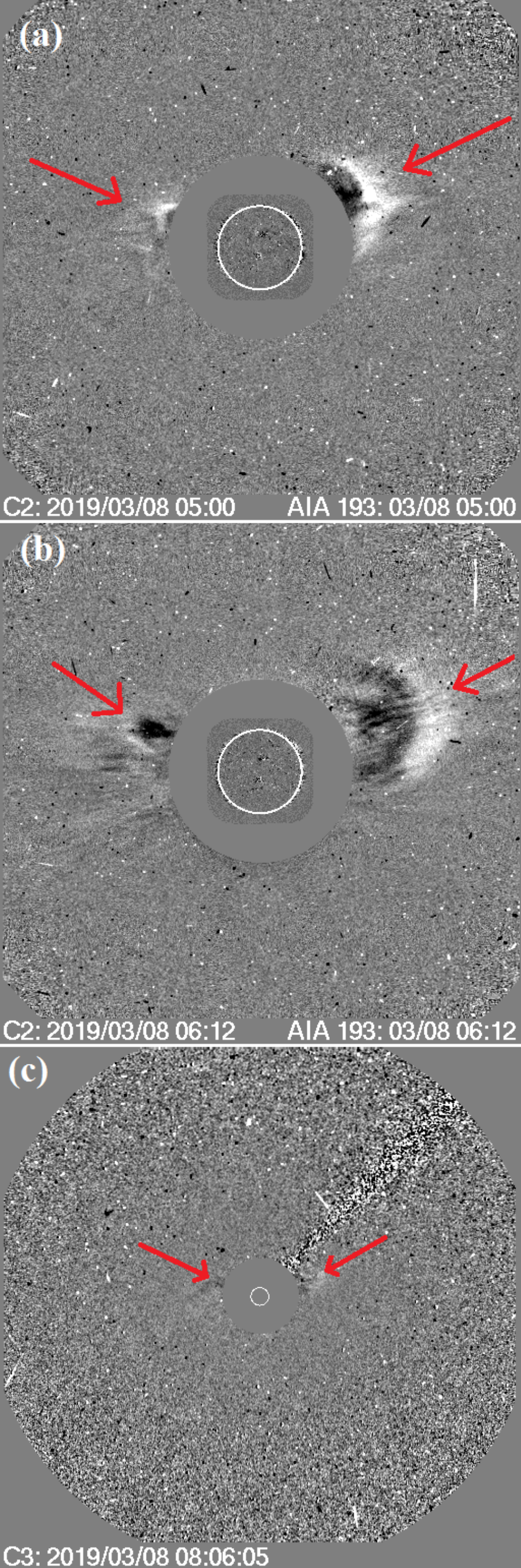}
\caption{LASCO C2 (\textbf{a} and \textbf{b}) and C3 (\textbf{c}) coronagraph running difference images showing the CME associated with the C-class eruptive flare on 8 March 2019 peaking at $\approx$03:30~UT. The CME is characterized by a weak event which faded soon after its appearance in C3 (\textbf{c}). Notably, the source region of the eruption lies close to the centre of the solar disk (see Figure~\ref{Fig_AR}) which resulted into a CME directed almost directly toward the observer. The arrows indicate the eastern and western signatures of the CME.}
\label{Fig_cme}
\end{figure*}

The successful eruption of the flux rope leads to a CME which was imaged by the LASCO coronagraphs (Figure \ref{Fig_cme}). The CME exhibited a poor visible appearance ({\it i.e.} reported as a weak event in LASCO CME catalogue) and became indistinguishable with the background sky soon after its detection in the LASCO C3 field of view (indicated by the arrows in Figure \ref{Fig_cme}).

\section{Radio observations by Udaipur-CALLISTO}
\label{sec:callisto}

\begin{figure*}
\centering
\includegraphics[width=\textwidth]{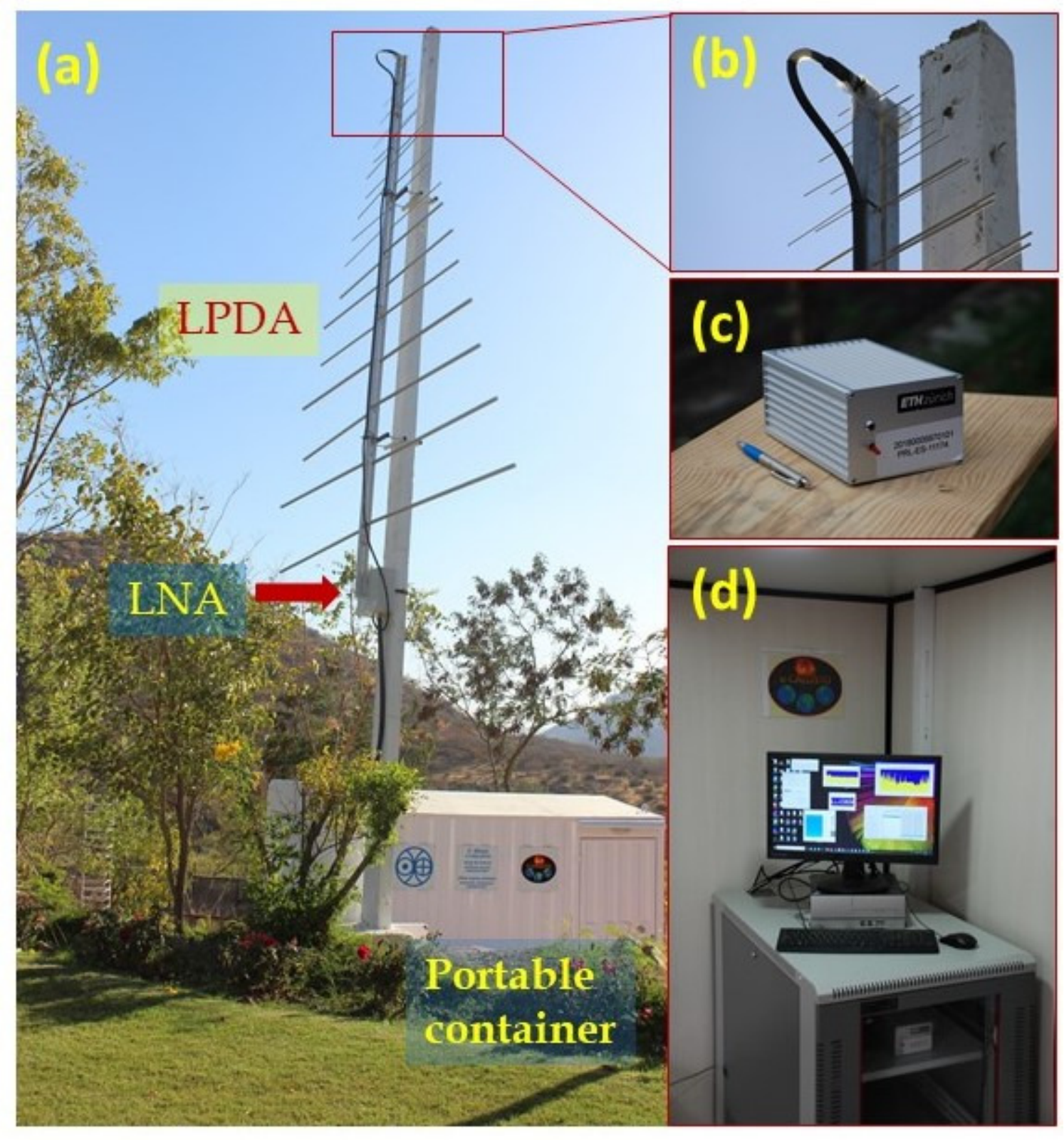}
\caption{Various subsystems of the Udaipur-CALLISTO: (\textbf{a}) shows the Log Periodic Dipole Antenna (LPDA) of height 3.63 m which is mounted toward zenith. A Low Noise Amplifier (LNA), marked by red arrow, is connected between the LPDA and CALLISTO spectrograph. (\textbf{b}) shows the zoomed image of the top portion of the antenna having smaller elements of LPDA and coaxial feed connection. A portable container houses the CALLISTO spectrograph (shown in (\textbf{c})) which is connected to a computer for data acquisition and storage (shown in (\textbf{d)}).}
\label{Fig_e-callisto_uso}
\end{figure*}

Udaipur Solar Observatory (USO) of Physical Research Laboratory (PRL) has recently commissioned a CALLISTO solar radio spectrometer\footnote{See https://www.prl.res.in/$\sim$ecallisto/ .} in Udaipur/India (24.61$^\circ$N 73.67$^\circ$E; altitude of $\approx$615 m above mean sea level) which is operational since 5 October 2018. The CALLISTO is essentially a \textit{Compound Astronomical Low cost Low frequency Instrument for Spectroscopy and Transportable Observatory} \citep[][]{Benz2009}. The instrument natively operates between 45 and 870 MHz with a frequency resolution of 62.5 KHz. An indigenously designed and developed Log Periodic Dipole Antenna (LPDA) has been deployed at Udaipur station, to collect the solar flux. The specification of LPDA is given in Table~\ref{Tab:callisto_summary}. The radio spectrometer attached to the antenna is a programmable heterodyne receiver developed at ETH Zurich, Switzerland. Different panels of Figure~\ref{Fig_e-callisto_uso} show various subsystems of Udaipur-CALLISTO that include LPDA (panels a and b), Low Noise Amplifier (LNA) (marked by red arrow in panel a), the CALLISTO spectrograph (panel c), computer for data acquisition and storage (panel d). A portable container is installed near the LPDA (seen in panel a) which houses the CALLISTO spectrograph, computer, and Uninterrupted Power Supplies (UPS). The UPS provides emergency power to all the electrical subsystems when the mains power fails. The technical overview of the Udaipur-CALLISTO set up is presented in \citet{Upadhyay2019}.

\begin{table}
\caption{Summary of specifications of the Log Periodic Dipole Antenna (LPDA) of Udaipur-CALLISTO.}
\begin{tabular}{ll}
\hline
Frequency range & 45$-$870~MHz\\
Gain & 8--9 dBi\\
Beam width & 90$-$110 degree\\
VSWR & $<$2  \\
Return loss & $<$ -10 dB \\
\hline
\multicolumn{2}{c}{Parameters for mechanical design}\\
\hline
Number of elements & 28\\
Material for boom and elements & Aluminium\\
Length of each boom & 3.63 m\\
Cross section of each boom & 4 cm $\times$ 4 cm\\
Spacing between the booms & 1 cm\\
Total stub length & 0.937 m\\
\hline
\end{tabular}
\label{Tab:callisto_summary}
\end{table}

\subsection{Evolution of the radio emission}
\begin{figure*}
\centering
\includegraphics[width=\textwidth]{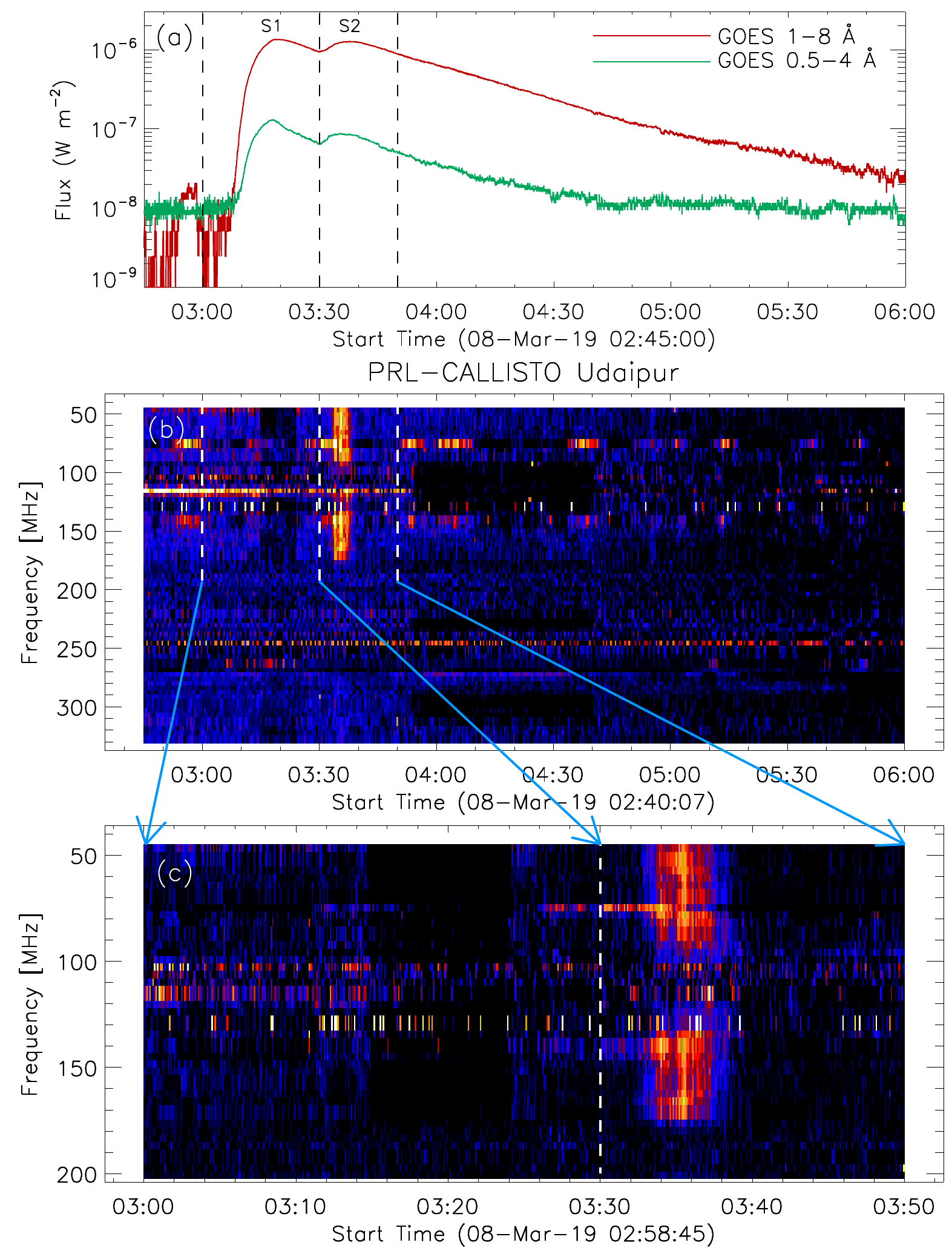}
\caption{Comparison of energy release during the long duration event in soft X-ray (observed from GOES) and low-frequency radio (observed from Udaipur-CALLISTO) emission. We note that ({\it cf.} (\textbf{a}) and (\textbf{b})) the radio emission is observed at $\approx$50--180 MHz during the stage II of the flaring activity (S2). (\textbf{c}) presents zoomed view of the CALLISTO dynamic radio spectrum. Note that the horizontal void at $\approx$90--125 MHz is due to the filtering of radio emission by the FM rejection filter and therefore should not be interpreted as the absence of solar radio emission at these frequencies.}
\label{Fig_ecallisto_1}
\end{figure*}

In Figure~\ref{Fig_ecallisto_1}, we provide a comparison between two-channel GOES SXR flux (panel a) and broad-band low frequency radio observations in the frequency range of 50--330~MHz observed by Udaipur-CALLISTO (panel b) during the flare. The dynamic radio spectrum clearly reveals flare associated radio burst between $\approx$50--180 MHz. The two stages of SXR flare emission are denoted by S1 and S2 in Figure~\ref{Fig_ecallisto_1}a. A comparison of top and middle panels of Figure \ref{Fig_ecallisto_1} readily shows the presence of radio emission only during the stage 2. For clarity, we further show zoomed image of the selected time and frequency range in Figure~\ref{Fig_ecallisto_1}c. It is worth mentioning that the horizontal void noticeable during $\approx$03:32--03:38~UT at $\approx$90--125~MHz is due to the filtering of radio emission by the FM rejection filter and therefore should not be interpreted as the absence of solar radio emission at this frequency range. 
The radio burst might have extended to lower frequencies, observations of which is beyond the lowest frequency limit of the instrument.

\subsection{Emission at low and high frequency bands}
\begin{figure*}
\centering
\includegraphics[width=\textwidth]{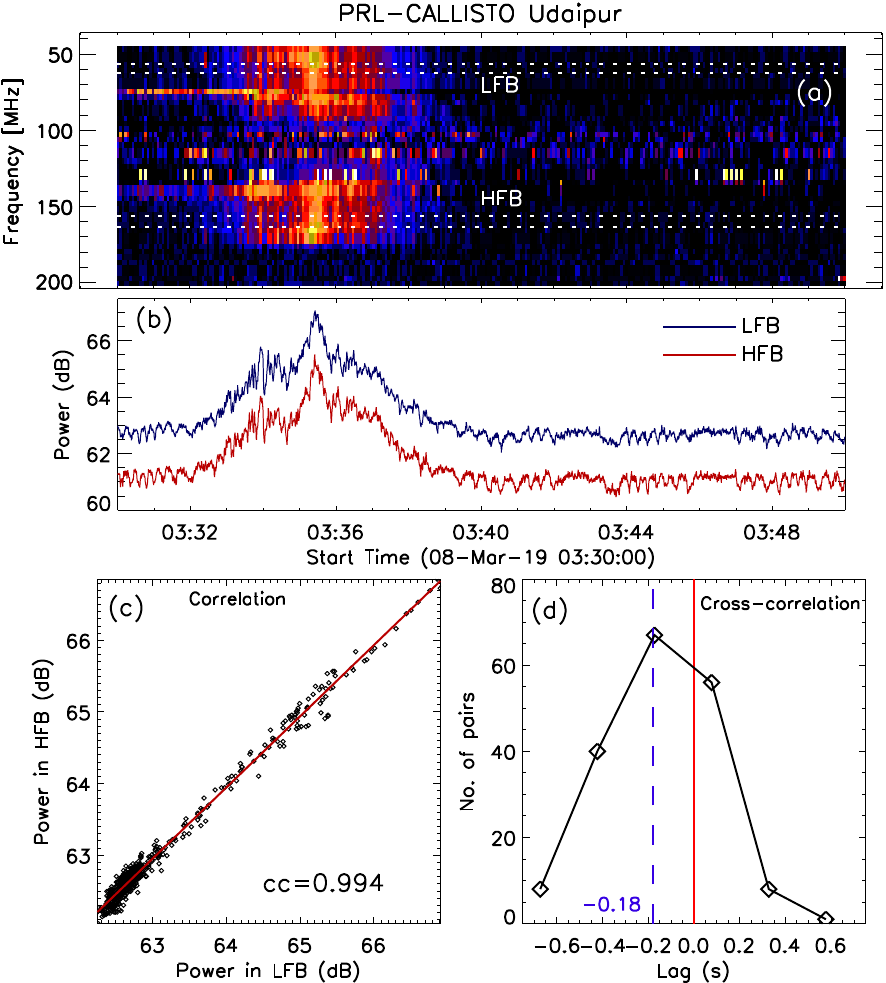}
\caption{(\textbf{a}): The CALLISTO dynamic radio spectrum focussing on the intensity and structures of radio emission at low-frequency band (56--62~MHz; LFB) and high-frequency band (156--164 MHz; HFB). (\textbf{b}): Flare light curves for LFB and HFB. (\textbf{c}): Correlation between radio emission at LFB and HFB showing the emission at two bands are highly correlated. (\textbf{d}): Cross-correlation between the radio emission at LFB and HFB calculated with a bin size of 0.25 s, showing a delay of 0.18 second.}
\label{Fig_ecallisto_2}
\end{figure*}

In order to understand the temporal evolution of the radio emission, we analyse the light curves at two frequency bands as shown in Figure~\ref{Fig_ecallisto_2}. The low-frequency band (LFB) ranges between $\approx$56--62~MHz while high-frequency band (HFB) ranges between $\approx$156--164~MHz (Figure~\ref{Fig_ecallisto_2}a). The time profiles in LFB and HFB suggest similar variations (Figure~\ref{Fig_ecallisto_2}b) with peak at $\approx$03:35 UT. We also notice a sub-peak prior to the overall maxima at about a minute earlier (\textit{i.e.} $\approx$03:34 UT). The correlation between the powers in LFB and HFB indicates excellent correspondence between them with a correlation coefficient of 0.994 (Figure \ref{Fig_ecallisto_2}c) which confirms our expectation that the emission at higher and lower frequency bands of the burst are highly synchronized.

In order to explore the frequency drift in the radio burst, we perform cross-correlation analysis of the time series of LFB (\textit{i.e.} $\approx$56--62 MHz) and HFB (\textit{i.e.} 156--164 MHz) in Figure \ref{Fig_ecallisto_2}d. We find a lag of 0.18 s. 
Notably, the observed time lag of 0.18 s for a frequency interval of $\approx$100 MHz is an order of magnitude less than that for a typical type III radio burst \citep{Alvarez1973}. 
In fact, considering the chosen bin size of 0.25 s, the cross-correlation results are consistent to effectively no drift between the two frequencies. 
The lack of any significant spectral drift in the radio emission suggests that the observed feature is stationary in nature, resembling a type IV burst.

\subsection{Composite Udaipur-CALLISTO and Wind/WAVES spectra}
\label{sec:callisto_UD_Wind}

In Figure~\ref{Fig_callisto_wind}, we compare radio observations from Udaipur--CALLISTO and Wind/WAVES. The dynamic radio spectrum from Wind/WAVES, covering a frequency range of 14 MHz$-$20~KHz, reveals a type III burst which starts at $\approx$03:28~UT at a frequency of $\approx$2.5~MHz. The composite spectra clearly indicate that the stationary type IV burst in the frequency range of $\approx$50$-$180~MHz observed by CALLISTO is delayed by $\approx$5~min with respect to the onset of type III burst measured at lower frequencies by Wind/WAVES. The starting frequency of type III radio burst at 2.5~MHz is consistent with a height of $\approx$4~$R_{\odot}$ \citep{Leblanc1998}.

\section{Coronal structure of active region}
\label{sec:extrapolation}
We employ the Non-Linear Force Free Field (NLFFF) modelling to understand the coronal magnetic field structures of the AR.  For the purpose, we selected an HMI magnetogram at 02:48~UT on 8 March 2019 which represents the photospheric configuration prior to the event. In Figure~\ref{Fig_extrapolation}, we show the results of NLFFF extrapolation in different viewing angles and field of views. From extrapolated field lines, we can consider two sets of field lines: inner and overlying. It is obvious that the flare ribbons essentially form at the footpoints of inner field lines. For clarity in our description, we have shown inner field lines by three colors: blue, green, and yellow. A comparison of model coronal field structure with AIA 304~\AA~image at $\approx$03:19 UT (during the first peak of the flaring event; Figure \ref{Fig_extrapolation}c) clearly suggests that the stage I of the flare predominantly involved the blue lines initially and then the green lines since the flare ribbons formed during this interval are exactly co-spatial with the footpoints of the blue and green lines 
({\it cf.} Figure~\ref{Fig_ribbons}). Further, the extrapolation results also clearly explains the structure and spatial extension of flare ribbon of stage I with extended northern ribbon and compact southern one (see Figure~\ref{Fig_ribbons} and Section~\ref{sec:loops_ribbons}). The flare ribbon of stage II essentially form at the footpoints of yellow (inner) field lines (Figure \ref{Fig_extrapolation}d). In Figure~\ref{Fig_extrapolation}a, we also show the overlying field lines associated with the AR by red color. As clearly seen, the overlying field lines represent the large-scale coronal loops which can be clearly seen in the side view of magnetic field extrapolation shown in Figure~\ref{Fig_extrapolation}b. 
In this context, we especially emphasize a positive polarity region at the north-eastern part of the AR which is denoted by a circle in Figure~\ref{Fig_extrapolation}a. We note this region to be the origin of large-scale overlying field lines (\textit{i.e.} red lines). This region also forms the footpoints of inner (closed) field lines which are drawn by yellow color in Figures~\ref{Fig_extrapolation}a--c.

In Figure~\ref{Fig_decay_index}, we graphically illustrate the variation of the magnetic decay index $n$ along the approximate PIL which generally forms the main axis of the coronal sigmoid. Magnetic decay index is expressed as $n=-\frac{\log(B_h)}{\log(z)}$, $B_h$ and $z$ being the horizontal component of magnetic field and height, respectively. \cite{Demoulin2010} have demonstrated that the critical value of the magnetic decay index for torus instability of the flux ropes \citep{Kliem2006} is $\approx$1.5. The approximate PIL is overplotted on magnetogram image in Figure~\ref{Fig_decay_index}a while the corresponding plots showing the decay index curve (i.e., variation of height verses length along the PIL for a given $n$) for the critical value of $n$=1.5 is shown Figure~\ref{Fig_decay_index}b. As such, the magnetic decay index is an important parameter to quantify of the strength of the magnetic confinement to a flux rope imposed by the overlying field lines of the active region \citep[\textit{e.g.}][]{Mitra2020b,JoshiNC2021,Mitra2021}. The Figure~\ref{Fig_decay_index} readily shows a diverse distribution of $n$ in the western and eastern parts of the sigmoid.

\section{Discussion}
\label{sec:discussion}
In this paper, we present a multi-wavelength analysis of C-class eruptive flaring activity on 8 March 2019 from NOAA 12734 which exhibit the characteristics of long duration event (LDE). This active region appeared in March 2019 when solar cycle 24 was running toward the end phase. Importantly, the paper presents the first observation of a solar flare by Udaipur-CALLISTO. GOES SXR profile in 1$-$8~\AA~channel clearly shows that the emission returned to pre-flare background level $\approx$3.5 hours after the onset of the flare. The prolonged soft X-ray emission imply that the hot plasma in the flaring region, impulsively heated during the magnetic reconnection, cooled down very gradually. 

\begin{figure*}
\centering
\includegraphics[width=\textwidth]{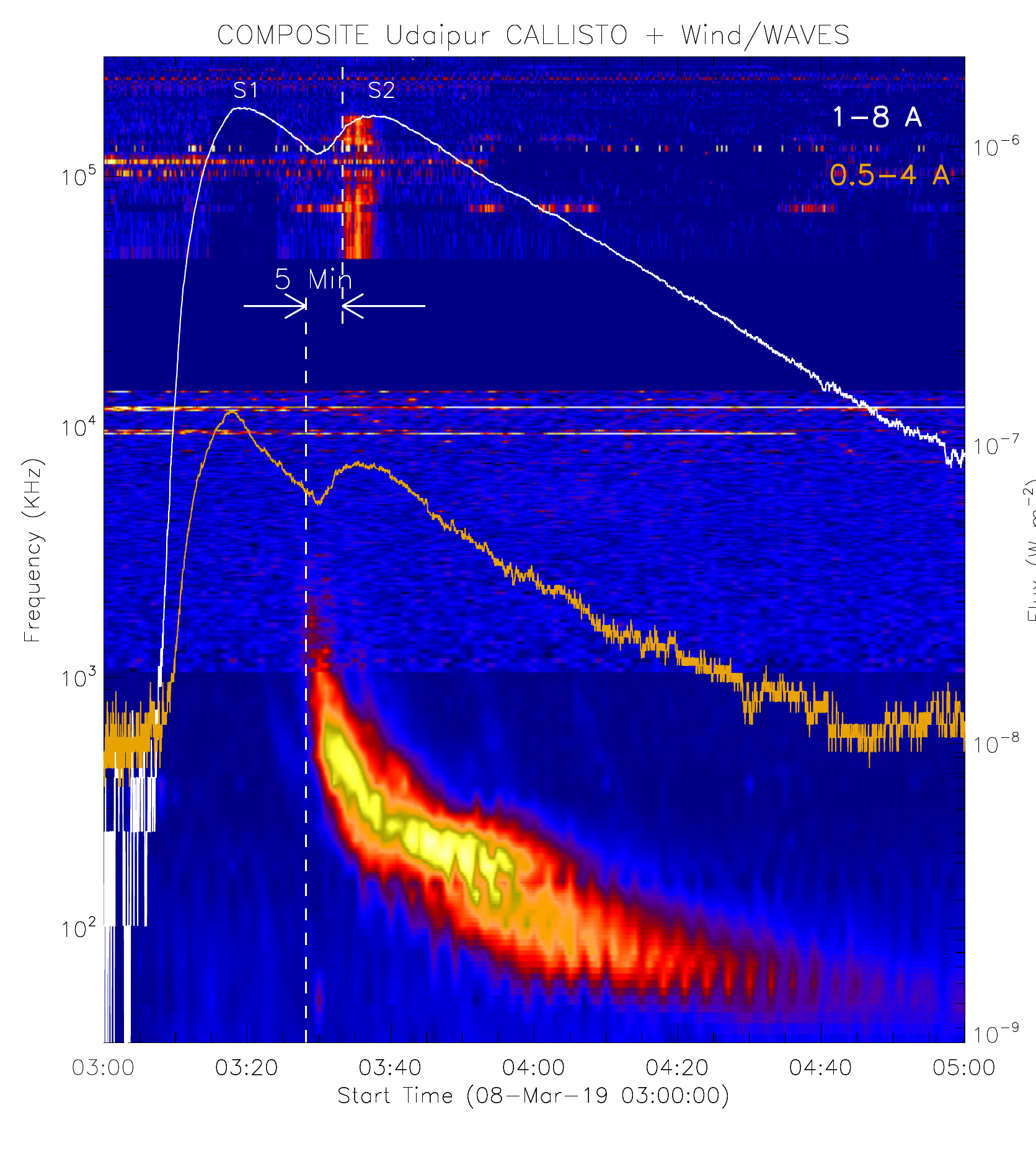}
\caption{(\textbf{a}): Composite dynamic radio spectrum from Udipur-CALLISTO and Wind/WAVES. The onset times of DH type III radio burst and stationary metric radio burst are marked by dashed lines at $\approx$03:28~UT and $\approx$03:33~UT, respectively, indicating a time difference of $\approx$5~min between the two bursts. We also note the absence of DH type III counterpart in metric wavelengths.}
\label{Fig_callisto_wind}
\end{figure*}

The coronal images of the active region NOAA 12734 in hot coronal channels of AIA (94 and 193~\AA) clearly reveal that the active region loops are twisted and form an overall S-shape, \textit{i.e.} a coronal sigmoid. Here it is worth noting that the sigmoidal active regions were originally identified in SXR images of the Sun by Yohkoh spacecraft \citep{Manoharan1996,Rust1996}. Later, it was recognized that sigmoids can also appear in EUV images, especially in hot channels ({\it e.g.} 94, 131, and 193~\AA~channels of AIA/\textit{SDO}), which implies abundance of structured hot thermal emission from twisted loops even during the quiet phase of the active region \citep[see, {\it e.g.},][]{LiuC2007,ChengX2014,Joshi2018,Mitra2020}. Notably, sigmoidal active regions show higher association with eruptive events over non-sigmoidal active regions \citep{Canfield1999,Glover2000}. 

The photospheric structure of the active region requires further attention. We note the active region to have a noticeable sunspot (or a very compact sunspot group) in its leading part while the trailing part is devoid of intense magnetic field regions to be visible as sunspot. However, the HMI magnetograms indicate that a large concentration of magnetic flux is still present in the trailing part of the AR with more dispersed distribution than in the leading one. We also keep in mind that the active region had already started to decay (see Section~\ref{sec:AR12734}) a couple of days prior to the day of reported activity with the successive disappearance of its trailing part. Notably, the decay of sunspots did not lead to a proportionate simplification in the coronal connections between the leading and trailing magnetic flux regions. In fact, the existence of coronal sigmoid in NOAA 12734 evidences that, although the AR poses a relatively simplified structure of photospheric magnetic fields, coronal loop system associated with it, was complex and contained excess energy to power an eruptive event.

\begin{figure*}
\centering
\includegraphics[width=\textwidth]{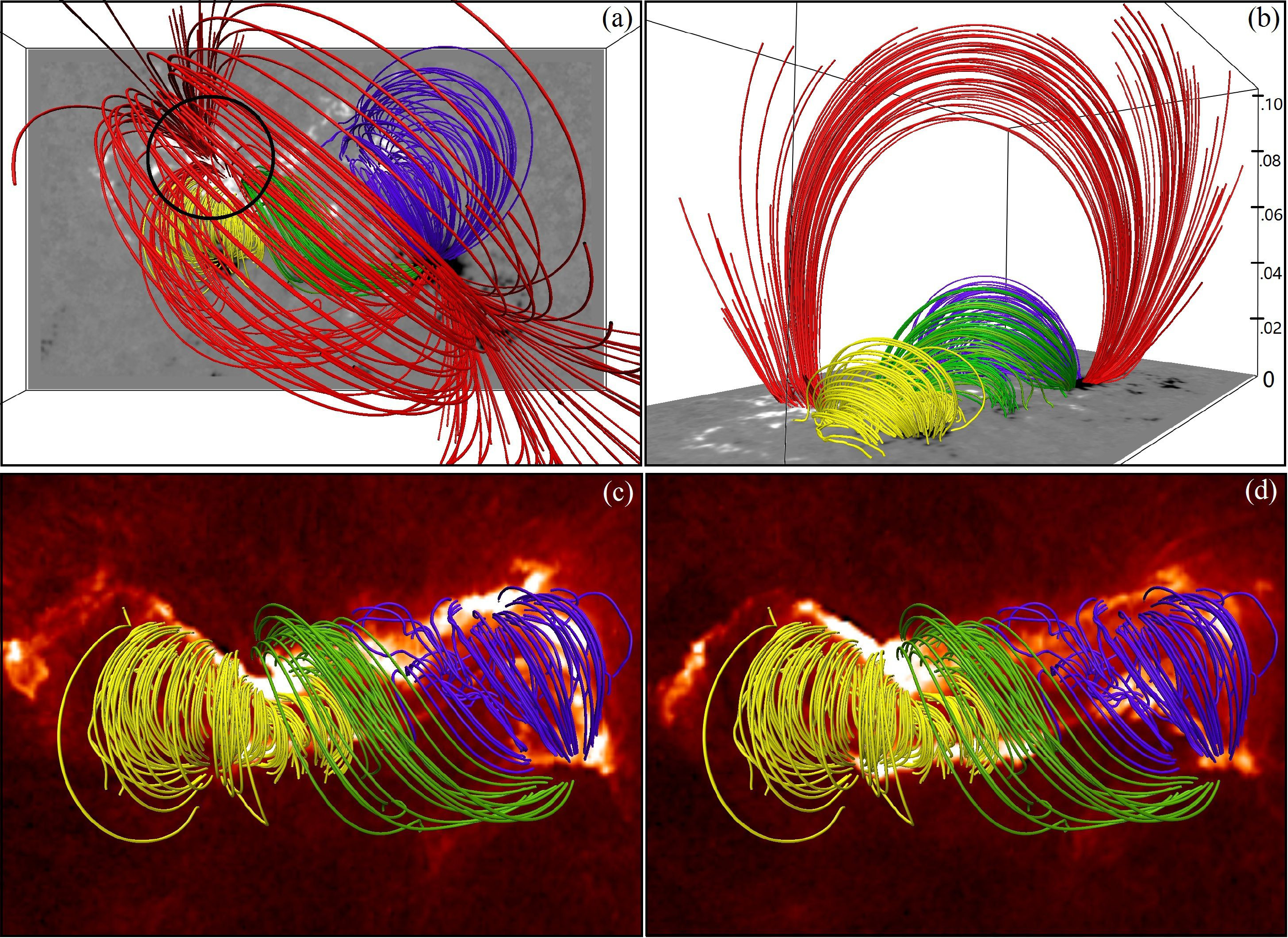}
\caption{NLFFF extrapolation showing model coronal magnetic field structures of the AR. We consider two sets of field lines: inner and overlying. The inner field lines are shown by blue, green, and yellow lines while the overlying field lines (in \textbf{a} and \textbf{b}) are delineated by red lines. \textbf{a} and \textbf{b} present top and side views, respectively. The levels in $z$-axis in (\textbf{b}) mark the height above the photosphere in the unit of solar radius. The bottom background in (\textbf{a}) and (\textbf{b}) are LOS magnetogram at 02:48~UT. In (\textbf{c}) and (\textbf{d}), we show only the inner field lines with AIA 304 \AA\ images at 03:19 UT (during the first peak) and 03:39 UT (during the second peak), respectively, as the background.}
\label{Fig_extrapolation}
\end{figure*}

The temporal evolution of the flare in SXR shows two successive stages of energy release with distinct peaks at an interval of $\approx$19 min (Figure~\ref{Fig_goes}). We notice that the light curves in both the GOES channels undergo a gradual decline after the first peak and reach a minimum level before the onset of the second stage of energy release which is suggestive of two distinct stages of magnetic reconnection. However, further analysis of AIA/\textit{SDO} images reveal that the dual-stage reconnection is associated with the eruption of a single pre-eruptive magnetic structure \textit{i.e.} a sigmoidal flux rope. The multi-wavelength investigations of the flaring region reveal several important aspects of the coronal energy release (Figures~\ref{Fig_fl_131_171}, \ref{Fig_arcade}, \ref{Fig_ribbons}) which are worth discussing. The site of energy release during the stage I of the flare is predominantly situated in the leading part of the active region. 
The first stage of reconnection forms hot coronal loop system (Figures~\ref{Fig_fl_131_171}c) besides relatively short flare ribbons (Figure~\ref{Fig_ribbons}a). Importantly, during the first stage, we also observe relatively faint yet clear enhancement of structured ribbon-like emission in the trailing part of the active region where the northern flare ribbons of the second stage later developed. Contextually, the NLFFF extrapolation results (Figures~\ref{Fig_extrapolation}) clearly show a set of field lines (drawn in green) that connects southern flare ribbon of stage I having compact structure, with northern flare ribbon formed in the trailing part of the active region. Thus, we find that both blue and green field lines are involved in the stage I of the flaring process. Subsequently, as evident from the apparent movement of flare ribbons and loops, the site of energy release shifts toward the eastern side of the AR.
Consequently, during the second stage of flaring process, the energy release dominates in the trailing part of the active region and proceeds with the formation of corresponding post-reconnection loop system (Figures~\ref{Fig_fl_131_171}e--h) and flare ribbons (Figure~\ref{Fig_ribbons}b). The development of large conjugate flare ribbons and growth of extended post-flare arcades spanning both the stages of the long-duration event show nice agreement with the standard model of solar flare \citep[\textit{e.g.} reviews by][]{Fletcher2011, Shibata2011, Joshi2012}. The ribbons of a two-ribbon flare essentially reflect the linkages of lower solar atmosphere with the energy release site in the corona through coronal magnetic field \citep[\textit{e.g.}][]{Warren2001, Chandra2009, Joshi2009, Joshi2017b} either by direct bombardment of electrons \citep{Masuda2001, Fletcher2001} or/and, thermal conduction \citep{Czaykowska1999, Czaykowska2001}.
By comparing NLFFF extrapolation results with the EUV observations, we find that  the stage II of the energy release is associated with yellow field lines shown in Figure~\ref{Fig_extrapolation}c--d. 
The successive occurrences of magnetic reconnection and resultant energy release in the decay phase are thought to be the cause of LDE flares \citep{Isobe2002}. 

The EUV images at multiple channels of AIA reveal the evolution of a system of coronal loop during the decay phase (Figure~\ref{Fig_arcade}). We can readily recognize the loop system to be very dense and extended in a large region that almost encompassed the entire AR. We further note that even after several minutes of the second peak of energy release, bright post-reconnection loop system of the stage I persisted (denoted as set I in Figure~\ref{Fig_arcade}). We attribute the total thermal emission during the decay phase (represented by GOES SXR flux) to be the composite of emission from post-flare arcades developed during the two-stage reconnection process (\textit{i.e.} composite of loops denoted as set I and set II in Figure~\ref{Fig_arcade}). 

\begin{figure*}
\centering
\includegraphics[width=\textwidth]{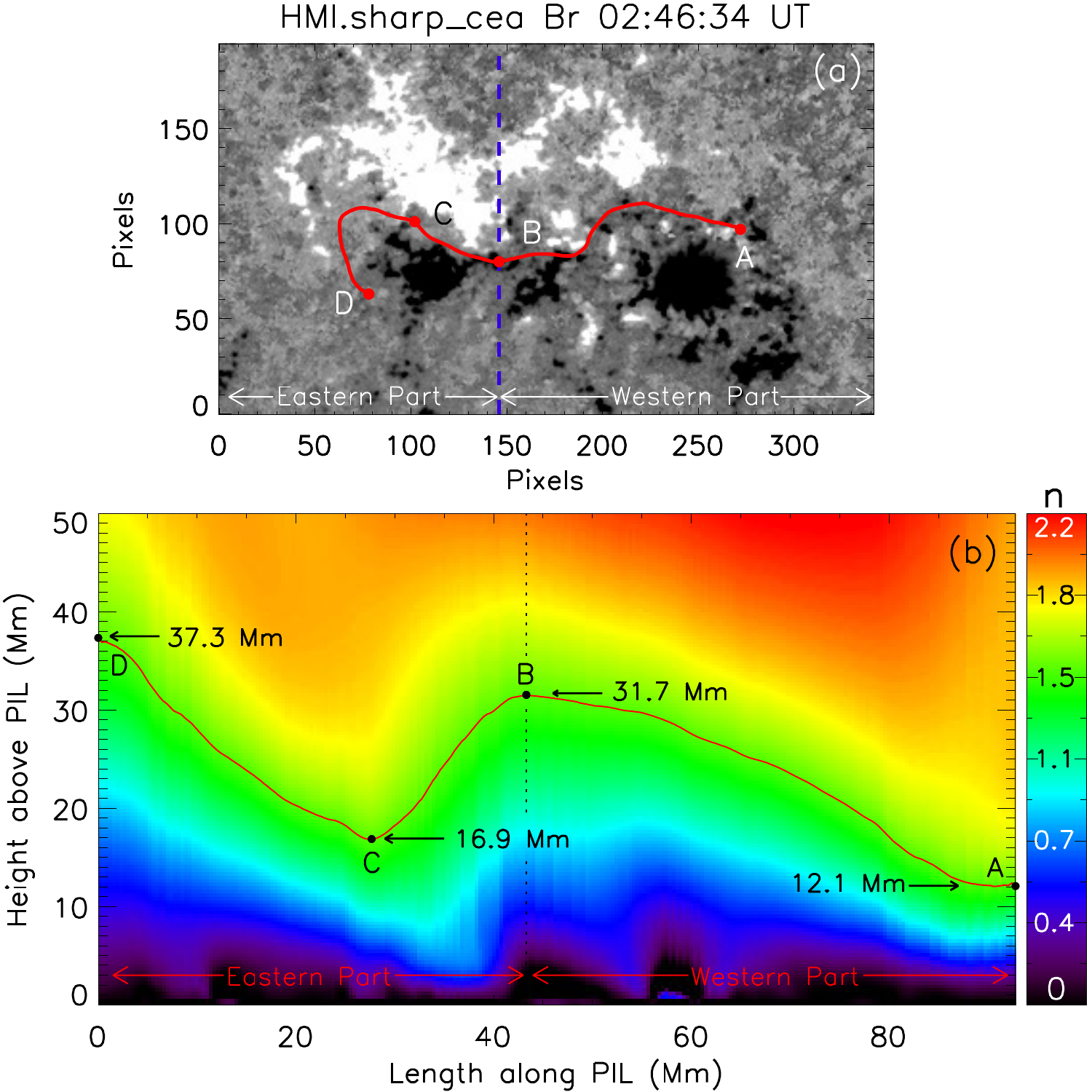}
\caption{(\textbf{a}): HMI LOS magnetogram showing the photospheric configuration of the AR NOAA 12734 prior to the flaring activity. The approximate polarity inversion line (PIL) is shown by red curve (ABCD). (\textbf{b}): Distribution of magnetic decay index ($n$) above the PIL.  The red curve refers to contour of n = 1.5.}
\label{Fig_decay_index}
\end{figure*}

The initiation of eruption at the western elbow of the sigmoid can be vividly understood in the light of distribution of magnetic decay index $n$ along the coronal sigmoid (Figure~\ref{Fig_decay_index}). The decay index curve clearly demonstrates that the confinement of the overlying magnetic field is the weakest at the western elbow of the sigmoid (i.e. location marked as `A' in Figure~\ref{Fig_decay_index}b) and becomes stronger toward the eastern part. However, beyond the central region of the sigmoid (indicated as region `B' in Figure~\ref{Fig_decay_index}b), the strength of overlying fields first decreases with height and then buids up again toward the eastern elbow. The synthesis of decay index curve together with the observations of sequential eastward shift of energy release site imply that the sigmoidal flux rope erupted asymmetrically under the influence of an external asymmetric magnetic confinement. Thus, the sigmoidal flux rope structure successively uprooted in a zipping-like motion from its western to eastern footpoints. The role of asymmetry in the spatial variation of overlying (\textit{i.e.} external) magnetic fields have been demonstrated in detail by \cite{LiuR2009} and our interpretation is well consistent with their proposal. Plausibly, the spatial variation in decay index curve with a peak at the central region (\textit{i.e.} region marked as `B' in Figure~\ref{Fig_decay_index}b) foresee that the zipping-like uprooting of the flux rope will show two steps in its kinematic evolution, corresponding to activation of the eastern and western parts. This is validated in the observations in the forms of time delay between the appearance of flare signatures (ribbons and post-flare arcade; Figure~\ref{Fig_arcade} and \ref{Fig_ribbons}) in the eastern and western part of the sigmoid together with twin EUV dimming regions (Figure~\ref{Fig_coronal_dimming}). The southwest and northeast dimmings (D1 and D2 respectively) are associated with the negative and positive magnetic polarity regions (Figure~\ref{Fig_coronal_dimming}f). Notably, the twin dimming regions observed here remain localized and stationary. In simplistic configuration, such twin-dimmings are recognized as the footpoints of the ejected flux rope which are rooted at the opposite magnetic polarities of the source region \citep[see][]{Hudson1996,Webb2000,Dissauer2018}.

\cite{Ning2018} conducted a comprehensive multi-wavelength study of a confined flare that exhibited two distinct stages of energy release as evidenced by X-ray time profiles at multiple energy bands. Being a near-limb event, they minutely examined the dynamic evolution of flare loop system during the two stages together with the X-ray and microwave spectra. Their analysis reveals that the two-stage evolution is associated with interaction between different system of loops with reconnection occurring at increasing altitude. Our analysis also shows two episodes of coronal energy release, however, the reconnection site in the corona mainly shifts laterally, parallel to the PIL. In the present case, slowly decaying emission after the flare peak is a manifestation of the extended post flare arcade which is essentially a composite structure developed during the dual-stage of magnetic reconnection associated with the sigmoidal flux rope eruption. While considering the total duration of elevated level of GOES flux, we should note that the reported event occurred near the end phase of the solar cycle when the overall activity level was extremely low with the pre-flare GOES 1--8~\AA~flux to be at $\approx$A0 magnitude (Figure~\ref{Fig_goes}). Obviously, the low background facilitates in the detection of weak flare emission. The much prolonged decay phase of this C-class event in SXR should be essentially viewed in terms of the lower threshold value of the background SXR flux originating from the solar disk.

The radio dynamic spectrum obtained from the Udaipur-CALLISTO spectrograph indicates that only the second stage of the flaring activity was associated with plasma emission at metric wavelengths (Figure \ref{Fig_ecallisto_1}). The intense plasma emission occurs without any appreciable frequency drift in the frequency range of $\approx$50--180 MHz that sustained for $\approx$7 minutes (Figure~\ref{Fig_ecallisto_2}. 
We attribute these observational findings as the signature of continuum emission, resembling a type IV burst \citep[\textit{e.g.}][]{Pick1986,Sasikumar2014}. In our case, the burst exhibits stationary nature, which is attributed to trapping of mildly energetic electrons in closed field lines during the post-reconnection regime \citep{LaBelle2003, LiuH2018}. The origin of radiation implies ongoing acceleration of electrons somewhere in newly formed arcades, possibly at the tops of the loops.
The coronal density model of \citet{Newkirk1961} suggests that the source of the observed burst is located between the height range of $\approx$1.1--1.5 $R_{\odot}$. It is worth mentioning that a type III burst observed by Wind/WAVES, starting at hectometre wavelengths ($\approx$2.5 MHz), precedes metric continuum emission (50--180 MHz) by $\approx$5 min (see Figure~\ref{Fig_callisto_wind}). The comparison of timings and frequencies of the two radio bursts at widely separated frequency domains provides, to some extent, a quantitative assessment about the onset times and formation heights of the two physical processes: the reconnection-opening of high-lying coronal loops triggered by the erupting flux rope and the subsequent formation of stable post-flare loop system after the complete ejection of the flux rope.

The comparison of spatial association of energy release site with the surrounding/overlying coronal magnetic field configuration by NLFFF modelling (Figure \ref{Fig_extrapolation}) provides important clues about the association of stationary plasma emission at metric wavelengths with stage II of the flaring activity. As noted earlier, the flare ribbons of stage II were associated with the inner field lines which are drawn in yellow (Figure \ref{Fig_extrapolation}). Here the common origin of northern footpoints of inner field lines of stage II and large-scale loops originating at positive polarity region (see the encircled region in Figure \ref{Fig_extrapolation}a) is noteworthy. 
We propose that with the progression of eruption, the large-scale loops of the AR (red ones shown in Figures \ref{Fig_extrapolation}a and b) get stretched and reconnected by the expanding flux rope; the eruption advances with the formation of a new set of large-scale loops at similar altitude. However, these newly formed large-scale loops undergo a continuous reorganization as the eruption of remaining part of the flux rope (rooted in the eastern side) is under way. Hence, the disruption of active region loops (both pre-existing and developed after post-reconnection) and formation of new ones go on simultaneously until the full structure of the asymmetrically activated flux rope expands beyond the height of large-scale loops connecting the northeast positive polarity region and southwest negative polarity region.  Only after the eruption of full sigmoidal structure (\textit{i.e.} when stage II is over), a stable large-scale loop system is created. The temporal association of the narrow-band metric radio burst with stage II of the energy release, along with the magnetic field configuration of the flaring region, suggests that the trapping of flare-accelerated electrons likely occurred near the apex of these post-reconnected large-scale, high-lying coronal loops. The continual restructuring of large-scale overlying loops during stage I likely prevents the trapping of electrons and corresponding radio signatures during this period.


\section{Summary and conclusion}
We have carried out a multi-wavelength analysis of an extended flaring activity, exhibiting a two-stage evolution, from a sigmoidal active region NOAA 12734. The paper also presents a description of the newly commissioned Udaipur-CALLISTO system of PRL which is a valuable addition to the existing programs of the Udaipur Solar Observatory and has expanded observing capabilities of the group to radio wavelengths. The multi-wavelength investigation reveals the flare to exhibit a dual-stage of energy release which is manifested both in terms of temporal and spatial evolution. The flare light-curves indicate two distinct peaks within an interval of $\approx$19 min. Importantly, as the second stage approaches, we note a successive lateral movement of brightness at both sides of the polarity inversion line 
from western part of the AR toward the east which eventually evolved into a large, ``classical" two-ribbon structures. Further, the spatial evolution of parallel ribbons follows the sequential development of the corresponding post-flare arcades.  We interpret apparent eastward shift in the formation of ribbons and loop arcades as evidence for the lateral progression of magnetic reconnection site as the sheared flux rope, traced by the sigmoid, activates and undergoes upward expansion. Our study also reveals that such lateral progression of reconnection site is actually driven by the asymmetrical eruption of the flux rope. We find that the western end of the flux rope activated first, causing the onset of magnetic reconnection as evidenced by the formation of flare ribbons in the region. It is followed by the successive uprooting of the eastern part of the flux rope and simultaneous development of the corresponding set of parallel ribbons. Thus we propose that the zipping-like asymmetrical eruption of the flux rope caused successive formation of eastern and western parts of the flare ribbons and overlying post-reconnection arcades. The observations also suggest that, in response to the overlying asymmetric magnetic field confinement, the flux rope on the whole undergoes two causally connected phases of fast kinematic evolution. The sequentially development of twin EUV dimmings that lie over the opposite magnetic polarity regions provide further credence to the above scenario; the twin-dimming regions being recognized as footpoint locations of the expanding flux rope. Notably, the successful eruption of the sigmoidal flux rope produced a CME, observed in the field of views of LASCO C2 and C3.

The morphological development of the flare inferred from EUV images are further complemented by coronal field modelling obtained from NLFFF method. In radio wavelengths, we observed a type III radio burst which starts at hectometre wavelengths ($\approx$2.5 MHz) which, after a few minutes, is followed by a stationary, broad-band ($\approx$50--180 MHz) radio emission, resembling a type IV burst. 
These radio observations underline the classical picture supporting reconnection-opening of field lines by flux rope eruption and subsequent formation of the stable post-flare arcade. 
Our observations also reveal that the asymmetrical eruption of the flux rope introduced noticeable delay in the activation of its two legs causing a prolonged phase of restructuring of overlying coronal loops. The continual changes in the topology of overlying field lines possibly leads to distinct difference in terms of origin of radio emission during the two stages. Based on the present work, we propose to carry out more such case studies of complex long duration events.


\begin{acks}
We are grateful to Dr. Anil Bhardwaj, Director, Physical Research Laboratory, Ahmedahad, India for his encouragement and support toward CALLISTO project at USO/PRL. We also sincerely thank Dr. Yashwant Gupta, Centre Director, NCRA-TIFR, Pune, India for providing technical expertise and facilities to fabricate the LPDA. We thank the \textit{SDO} and GOES teams for their open data policy. \textit{SDO} is NASA's missions under the Living With a Star (LWS) program. We also thank FHNW, Institute for Data Science in Brugg/Windisch, Switzerland for hosting the e-Callisto network. DO acknowledges support of the Department of Atomic Energy, Government of India, under the project no. 12-R\&D-TFR-5.02-0700. We thank Dr. Thomas Wiegelmann for providing the NLFFF code. We are also thankful to Binal Patel for help in the analysis of the radio spectrum. We are grateful to the referee of the paper for providing us with a very constructive set of comments and suggestions that enhanced the scientific content and presentation of the paper.

\noindent
{\bf Disclosure of Potential Conflict of Interest} The authors declare that they have no conflict of interest.
\end{acks}

\bibliographystyle{spr-mp-sola}


\end{article}
\end{document}